\documentclass{article}
\usepackage{hyperref}
\usepackage{graphicx}
\usepackage[top=2cm,bottom=2cm,left=2cm,right=2cm]{geometry}
\graphicspath{{figs/}}
\usepackage{wrapfig}
\usepackage{listings}
\usepackage{numberedblock}
\usepackage{placeins}
\usepackage{caption}
\usepackage{xcolor}
\usepackage{amsmath}
\usepackage[utf8]{inputenc}

\usepackage[english]{babel}

\usepackage{cite}

\DeclareCaptionFont{white}{\color{white}}
\DeclareCaptionFormat{listing}{%
	\parbox{\textwidth}{\colorbox{gray}{\parbox{\textwidth}{#1#2#3}}\vskip-4pt}}
\begin{document}

\bibliographystyle{plain}

\belowcaptionskip=-10pt

\title{Market-making with reinforcement-learning (SAC)}
\author{Alexey Bakshaev \\ alex.bakshaev@gmail.com }

\maketitle

\begin{abstract}
The paper explores the application of a continuous action space soft actor-critic (SAC) reinforcement learning model to the area of automated market-making. The reinforcement learning agent receives a simulated flow of client trades, thus accruing a position in an asset, and learns to offset this risk by either hedging at simulated "exchange" spreads or by attracting an offsetting client flow by changing offered client spreads (skewing the offered prices). The question of learning minimum spreads that compensate for the risk of taking the position is being investigated. Finally, the agent is posed with a problem of learning to hedge a blended client trade flow resulting from independent price processes (a "portfolio" position). The position penalty method is introduced to improve the convergence. An Open-AI gym-compatible hedge environment is introduced and the Open AI SAC baseline RL engine is being used as a learning baseline.
  
\end{abstract}

\section{Introduction}
Let's assume that our goal is to train our agent in a way that it can perform market making effectively. In this trading mode our agent puts out both the price it is willing its clients to buy at ("client" $ask$) and sell at ("client" $bid$), thereby accepting incoming flow of client orders at those prices and profiting from the resulting spread ($ask-bid > 0$). Client order flow would depend on a number of factors (explained in Environment section) making it non-symmetric, which results in our agent accruing a variable position in a traded asset with time. As the market price of the asset changes with time, this results in revaluation of the position, introducing uncertainty about the price the position would be closed at and hence the final profitability of transactions made over the lifetime of this position. \\
To give an intuitive example of this, consider Figure 1 which displays the dynamics of market mid price as well as bid and ask prices our agent is making in the market. 

\begin{wrapfigure}[14]{l}{0.5\textwidth}
	\caption{Example of unmanaged market risk}
	\centering
	\includegraphics[width=0.5\textwidth]{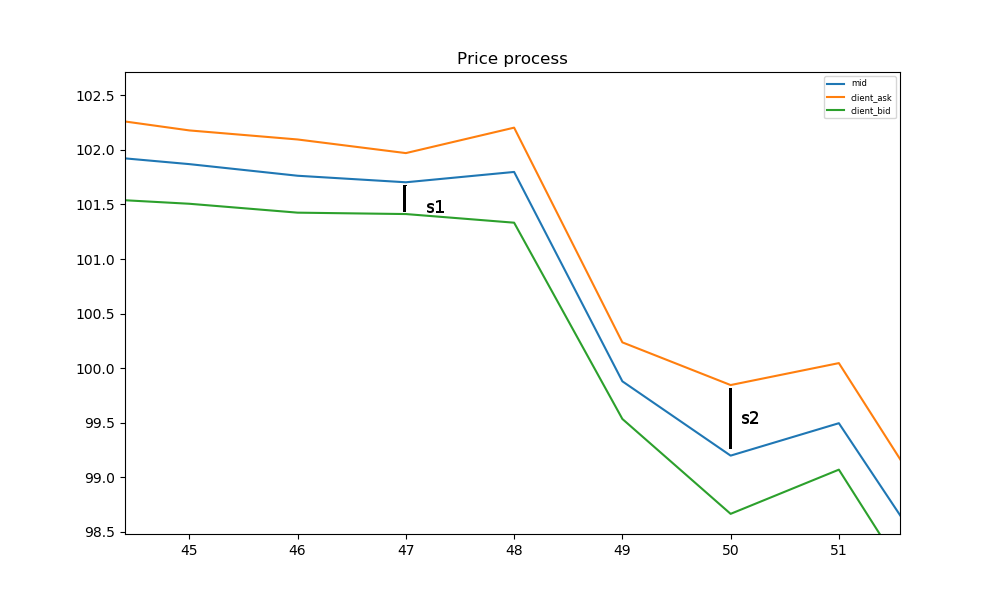}
\end{wrapfigure} 
At a time $t=47$ the agent buys from a client at a discount to mid market price (at "client bid"), makes a half-spread $s1$ and takes on a long position in an asset. If the agent was to get out of this position instantaneously at the same time $t=47$ by selling to another client, it would have made the entire spread of $ask-bid$. \\ 
Instead, however, the agent closes the position at time $t=50$  at a lower market at a half-spread $s2$, and the sum of those half-spreads $s1+s2$ is not sufficient to offset the drop in mid market price that causes the negative revaluation of the position.
\\\\
This raises the need in risk-managing which our hedging agent can achieve by:
\begin{itemize}
\item putting offsetting orders out to exchange and hereby "hedging" the position at the cost of paying exchange bid/ask spreads
\item decreasing client bid/ask spreads to attract more client flow on a chosen side and hereby decreasing the position more quickly. This is known as "skewing" the price under condition of client flow ("demand") being sufficiently elastic to price changes.
\end{itemize}
Hedging comes at a cost of decreased profitability whereas not hedging introduces the risk of incurring losses in case of unfavorable price movements. Thus, given the net incoming client order flow and the resulting accrued client position, the agent needs to learn to reason out its risk appetite: how much of the position to carry (in a hope it gets offset by the future client order flow) and how much of it to hedge either via an exchange or by offering clients more attractive prices to attract more offsetting client flow. \\
In a driftless environment where the asset price is a martingale, the expectation of the market price move with time is zero:
\begin{equation}
\label{martingale}
E[S_{t_2} - S_{t_1}] = 0 \;\;\; (t_2 \ge t_1 \ge 0)
\end{equation}
From this follows the first challenge in setting up an effective market-making ML framework: the agent needs to learn that there is no sense in entering speculative hedge positions that would try to profit from market moves. Instead, in a very basic setting, the agent needs to learn to offset the accrued client position with an opposite exchange (hedger) position. This is an inventory management problem that is very similar to a popular \href{https://gym.openai.com/envs/InvertedPendulum-v2}{Pendulum environment} where ML agent learns to balance the pendulum, or to \href{https://gym.openai.com/envs/MountainCar-v0}{Mountain car environment} where the agent needs to learn to swing the car out of the bottom of the pit.\\
Both observation space (the size of a position accrued by the agent) and action space (choice of an amount to hedge) are chosen to be continuous, and the asset price is driven by the drift-adjusted Brownian motion. This allows for an infinite action search space and presents a challenging task. An extension of this problem is presented when price skew is added as an additional action available for an agent to take making the overall search space a box in $R^2$.\\
Assuming the agent has successfully learnt  to balance the position to zero via choosing hedge / skew actions, we can now pose the following problem. In an environment where hedge spreads and client flow are unknown, can the market making agent learn to offer such spreads that would compensate it for the risk taking? In other words, can an agent learn the price of market risk, measured in terms of offered client spreads?\\
Another problem that naturally follows is that of portfolio management. The dynamics of the portfolio value is determined by the blended asset price process as well as the blended client trade flow, both of which are driven by the underlying asset price processes, correlation coefficients and asset weights. The learning agent does not know what correlations and asset weights have been used to build the portfolio and needs to work out the effective hedge strategy that will maximize profitablity and minimize the position involved. This makes it an inventory management problem with not just one asset but a portfolio of assets.

\section{Reinforcement learning and soft actor-critic (SAC)}
Reinforcement learning is becoming increasingly popular in the area of robotics and control automation. In the reinforcement learning setting there is a learning agent that can perpetually ($i \in [1, N]$) interact with the environment by performing an  $action_i$ and then observing the resulting $state_i$ of the environment  and the associated reward $r_i$. The main purpose of the reward function is to define the end goal of the learning process by rewarding "good" actions and penalizing for "bad" ones. As a result, the agent learns how to sample from the action space in a way that maximizes cumulative discounted return over all steps within a simulation trajectory
\begin{equation}
\label{Simulation trajectory reward}
R(\tau)=\sum_{i}^{}\gamma^i r_i
\end{equation}
where $\gamma$ is a discount factor which puts more weight on nearest rewards. The learned probability distribution $\pi $ for action sampling given reward increment and observation space state is called a $policy$. The goal of the learning process is to maximize the expected return over all simulation trajectories and to converge onto the optimal policy $\pi^*$:
\begin{equation}
\label{Max policy}
\pi^* = \arg\max_{\pi} E_{\tau \sim \pi}\left[ \sum_{i}^{}\gamma^i r_i \right]
\end{equation}
If an agent chooses actions according to some fixed policy $\pi$ starting from a given initial pair of state and action, the resulting expected return is known as "value function", representing the value of this policy. Naturally, as the network learns, the policy gets updated, resulting in a new value function on each step of a trajectory. A Q-function, or "action-value" function is a value function where the first action is being fixed (e.g. sampled off-policy) before the policy is applied on consecutive steps of the trajectory.\\
Soft actor-critic (SAC) algorithm introduced in \cite{haarnoja2018learning}, \cite{haarnoja2018soft} and \cite{haarnoja2018soft1} offered a few methods of improving stability of convergence to the optimal policy in case of continuous action and observation spaces:
\begin{itemize}
	\item \textbf{entropy maximization}: entropy $H(\pi(s_t))$ of the action-generating distribution is being used as a part of the reward function. This leads to joined maximization of expected policy and return which encourages stochastic exploration and stability of convergence:
	\begin{equation}
	\label{Max policy with entropy}
	\pi^* = \arg\max_{\pi} E_{\tau \sim \pi}\left[ \sum_{i}^{}\gamma^i (R(s_i,a_i,s_{i+1}) + \alpha H(\pi(s_t))) \right]
	\end{equation}
	where $\alpha$ is the temperature parameter which determines the trade-off between maximizing return vs maximizing entropy (exploration)
	\item \textbf{actor-critic}: the use of separate policy ("actor") and value function ("critic") networks. The policy $\pi$ and two value ($Q(s,a)$) functions $j=1,2$ are being concurrently learned so as to minimize the respective loss functions
	\begin{equation}
	\label{SAC Loss function}
	L_j=E\left[(Q_j(s_i, a_i)-y(r_i,s_{i+1}))^2\right]
	\end{equation}
	where the target $y$ is approximated by sampling actions ${\tilde{a}}_{i+1}$ from the policy $\pi(s_{i+1})$, conditioned on state return $r$ and next state $s_{i+1}$ coming from the replay buffer:
	\begin{equation}
	\label{SAC target}
	y(r, s_{i+1}) = r_i + \gamma(1-d)\left(\min_{j=1,2}Q^{targ}_j(s_{i+1},{\tilde{a}}_{i+1})-\alpha \log \pi({\tilde{a}}_{i+1}|s_{i+1}))\right), \;\; {\tilde{a}}_{i+1} \sim \pi(s_{i+1}) 
	\end{equation}
	The target value function ($Q_j^{targ}$) networks are being periodically updated from current value networks $Q_j$ in an exponential average fashion. Within the learning loop (see \nameref{SAC learning loop}) Q-functions are updated by gradient descent on the loss function \ref{SAC Loss function} and the policy is updated by gradient ascent on the entropy-adjusted Q functions.
	\item \textbf{experience replay}: the use of the replay buffer to simulate next states in an off-policy manner. Next actions are simulated from the current policy.
\end{itemize}

These methods alongside with others were used for baseline implementation of SAC algorithm   by \href{https://spinningup.openai.com/en/latest/algorithms/sac.html}{OpenAI}. It serves as a baseline RL engine used in this paper. The boxed network is a multi-layered ReLU perceptron with a linear input and output layers. Linear input layer transforms dimensionality from input (observation) space onto the first perceptron layer. Linear output layer transforms dimensionality from the last perceptron layer onto the action space. E.g. when we have the action space defined as position in an asset (dim = 1), and action space defined as hedge amount (dim = 1) and 2 ReLU layers of 2048 units, we will write this as MLP Linear (1) x ReLU (2048) x ReLU (2048) x Linear (1). An additional dimension to all those would be the batch size.

\section{Environment} 
Hedge environment is implemented to be \href{https://gym.openai.com}{Open AI Gym}-compatible. It makes use of Pytorch \href{https://pytorch.org/docs/stable/tensors.html}{Tensor} framework for generating price processes and client flows and is intended to be used with a Pytorch-based implementation. Hedge environment provides the standard members \texttt{step()} for updating the hedger state based on the action selected by the agent, \texttt{reset()} for re-setting the simulation and \texttt{render()} for displaying the dashboard with hedger statistics. Visualization of the hedging agent performance serves as a very important tool in understanding the progress (or lack thereof) of the hedging agent as well as assessing the overall correctness of the environment set-up.\\
Project code for this paper is available under: \textbf{https://github.com/bakalex/autohedger}

\subsection{Market data}
Asset price process is generated as a standard Euler discretization of a log-normal price process:
\begin{equation}
\label{Asset price process 1}
\Delta{log(S_i)}=\mu \Delta{t} - \frac{1}{2}{\sigma}^2 \Delta{t} + {\sigma}{\epsilon}\sqrt{\Delta{t}}
\end{equation}
\begin{equation}
\label{Mid price process}
S_N = S_0 \exp{\sum_i{\Delta{log(S_i)}}}
\end{equation}
For the inventory management problem we are trying to solve we don't want our reinforcement learning agent to focus on market drift which would result in the hedger taking speculative directional positions in an asset. Instead, we want it to focus on managing the position arising from the incoming client order flow, so we assume no drift ($\mu=0$) which makes our price process a martingale (eq \ref{martingale}).\\
For a given normally distributed variable $\epsilon_1$ and uncorrelated i.i.d variable $\epsilon$ a correlated normally distributed variable $\epsilon_2$ with correlation coefficient $\rho$ is generated as follows:
\begin{equation}
\label{Correlated}
\varepsilon_2 = f(\varepsilon_1, \rho) = \rho \varepsilon_1 + (\sqrt{1-{\rho}^2}) \varepsilon
\end{equation}
Generation of \href{https://pytorch.org/docs/stable/distributions.html#normal}{normally} distributed variables as well as above discretization is done within Pytorch Tensor framework so that at the beginning of the learning process we have the market data set ready for the entire simulation.
\subsection{Stochastic spread model}
\label{Stochastic spread model}
There are two types of offered spreads $\delta^{bid}$, $\delta^{ask}$ defined in the model: "client" spreads that our market-making agent offers clients to trade at (hence, "client" bid / ask) and "hedge" spreads that our agent trades at with an exchange (hence, "hedge" bid/ask). Both those spreads are applied on the top of the same mid price process \ref{Mid price process}. Generated spreads are based on the rolling mean volatility $\sigma_{roll\_avg}$ of the mid price process plus log-normal stochastic spread add-on $\varepsilon^L$.
\begin{equation}
\label{Bid price}
S^{bid} = S - \nu \delta^{bid}, \;\; S^{ask} = S + \nu \delta^{ask}
\end{equation}
\begin{equation}
\label{Stoch spread bid}
\delta^{bid}=\sigma_{roll\_avg}+\varepsilon^L_{1}, \;\; \delta^{ask}=\sigma_{roll\_avg}+\varepsilon^L_{2}
\end{equation}
\begin{equation}
\label{Rolling vol}
\sigma_{roll\_avg} = \frac{StDev(S_i, window)}{\sqrt{n\_steps}} 
\end{equation}
\begin{equation}
\label{Stoch spread ask}
\varepsilon^L = LogNormal\left(0, \gamma S_0 \frac{\sigma}{\sqrt{n\_steps}}\right)
\end{equation}
where $\nu$ is spread multiplier determining the overall magnitude of the spread, $\gamma$ is spread multiplier that determines the magnitude of the stochastic spread add-on, $\sigma$ is flat volatility driving the mid process and $\varepsilon^L_{1}$ and $\varepsilon^L_{2}$ are log-normal i.i.d variables. \\
Given the fat-tailed nature of the log-normal distribution, stochastic spreads $\delta^{bid}$, $\delta^{ask}$ are being clamped between 0.1 and 2.5 of mean simulation volatility. This helps to avoid unreasonable spikes and bid/ask prices.

\subsection{Client trade flow model}
\label{TradeFlowModel}
Client trade sizes hitting our offered bid and ask prices are simulated as bid and ask \href{https://pytorch.org/docs/stable/distributions.html#torch.distributions.poisson.Poisson}{Poisson} processes. Magnitude of intensity of those processes is assumed to be a function of rolling volatility of the price process. Net trade flow that determines how position changes on each step of the simulation is a function of imbalance of bid/ask \href{https://pytorch.org/docs/stable/distributions.html#torch.distributions.poisson.Poisson}{Poisson} intensities which is assumed to be correlated with the log-price process.
\begin{equation}
\label{Net trade size for a step}
TradeSize^{net} = TradeSize^{bid} - TradeSize^{ask}  
\end{equation}
\begin{equation}
\label{bid trade size}
TradeSize^{bid} = Poisson(ClientTradeRate * \bar{\lambda}^{bid})
\end{equation}
\begin{equation}
\label{ask trade size}
TradeSize^{ask} = Poisson(ClientTradeRate * \bar{\lambda}^{ask})
\end{equation}
where $ClientTradeRate$ determines the overall magnitude of the size Poisson process, and $\bar{\lambda}^{bid}$ and $\bar{\lambda}^{ask}$ are trade flow intensity multipliers. 
\begin{equation}
\label{Client trade rate}
ClientTradeRate = C \left(\alpha + \frac{{\sigma}_{rolling}}{{\sigma}_{mean}}\right)
\end{equation}
where $\alpha$ allows to define the vol-independent "mean" client trade flow, ${\sigma}_{rolling}$ is the average volatility within a rolling window and ${\sigma}_{mean}$ is mean volatility for the entire simulation. $C$ is the flow scaling factor and may be taken to be equal to the initial price of the asset $S_0$.\\Trade flow intensity multipliers are defined as:
\begin{equation}
\label{bid intensity}
\bar{\lambda}^{bid}=\max{(1-\lambda, 0)}
\end{equation}
\begin{equation}
\label{ask intensity}
\bar{\lambda}^{ask}=\max{(1+\lambda, 0)}
\end{equation}
where $\lambda$ is the net intensity. It is defined to be correlated with the log-price process as per eq \ref{Correlated}: \begin{equation}
\label{net intensity}
\lambda=\beta * \varepsilon_2 (\log{S_i}, \rho)
\end{equation} 
where $\beta$ is sensitivity of net client trade flow towards  log-returns and $\rho$ defines correlation between net intensity process and log-price process. To smoothen out variance in intensity, a rolling average version of the log-price process may be used.

\subsection{Profit and loss (PNL) model} 
As explained in introduction, our hedging agent accepts client trades at "client" spreads $\delta^{client}$ it offers to clients and then it may choose to hedge the accrued position out exchange "hedge" spreads $\delta^{hedge}$. Let's assume a client sells to our agent at our offered ask price and then the agent hedges out (sells) the accrued position to the exchange at exchange bid price. In a hypothetical setting where our hedger could have hedged instantaneously, it could have made the client half-spread and paid exchange half-spread, locking in the profit.
\begin{equation}
\label{Simple profit}
PNL = TradeSize (S^{client\_ask} - S^{mid}) - TradeSize (S^{hedge\_bid} - S^{mid}) = \left( \delta^{client\_ask} - \delta^{hedge\_bid} \right) TradeSize
\end{equation} 
In reality, having accrued the client trade at step $t_1$, our hedger can only choose to hedge out on the next step $t_2$ of the simulation, or even later, which results in revaluation of the position at the market mid price between those steps.
\begin{equation}
\label{Simple profit with market move 1}
PNL = TradeSize (S^{client\_ask}_{t_1} - S^{mid}_{t_1}) + TradeSize (S^{mid}_{t_2} - S^{mid}_{t_1}) - TradeSize (S^{hedge\_bid}_{t_2} - S^{mid}_{t_2})
\end{equation}
\begin{equation}
\label{Simple profit with market move 2}
= \left( \delta^{client\_ask}_{t_1} - \delta^{hedge\_bid}_{t_2} + \Delta{S^{mid}_{t_2-t_1}}  \right) TradeSize
\end{equation}
Given the martingale nature of the simulated price process $E \left( \Delta{S^{mid}_{t_2-t_1}} \right) = 0$ our agent can only be profitable if average offered client spreads are greater than average paid hedge spreads. Note that our reinforcement agent does not know that the price process is a martingale, so as it begins to explore the action space initially, it would be sensitive to realized directional market moves of the mid process. With time, however, the agent needs to learn to focus on managing positions resulting from the client flow instead of trying to capture directional moves of the mid process. Martingale formulation of the price process helps to achieve that. 

\subsection{Elasticity of demand (skewing) model}
\label{Skewing model explained}
When demand is elastic, we may affect the trade size a client is willing to trade with us by offering a more attractive ("skewed") price than the rest of the market. This can be used as means of position management: e.g. given a long position in an asset, we could lower our offered price to entice the client to buy from us thus decreasing the position. Depending on the marginal profitability of the transaction given the skewed price, such an action may be less expensive compared to executing the similar transaction on the exchange. We will be using the linear model in our experiments:
\begin{equation}
\label{Skewing model}
\Delta{size} = skew * \beta  \frac{MaxHedgeSize}{ClientTradeRate}
\end{equation}
where $ClientTradeRate$ is the mean trade rate from TradeFlowModel $\ref{TradeFlowModel}$. $MaxHedgeSize$ is the maximum hedge trade size constant defined for the bounded action space for the hedger and $skew$ is a bounded action $skew \in [-1.0,1.0]$ resulting in the following client price adjustment:
\begin{equation}
\label{Skewed prices bid}
S_{adj}^{bid}=S^{bid}-skew \left(S^{mid} - S^{bid}\right), \;\;skew \in [-1.0,0.0]	
\end{equation}
\begin{equation}
\label{Skewed prices ask}
S_{adj}^{ask}=S^{ask}-skew \left(S^{ask} - S^{mid}\right), \;\;skew \in (0.0,1.0]	
\end{equation}
So, negative skew means our agent buys more expensively and attracts more client flow on bid side. Positive skew means our agent sells cheaper and attracts more client flow on ask side.\\
$\beta$ constant is specified to be large enough for the learning agent to prefer skewing over hedging. Setting $\beta = 0$ makes the client flow non-elastic and can be used for price discovery purposes such as finding market price of risk spreads in $\nameref{Market price of risk}$. 

\subsection{Dummy hedger and simulation dashboard}
\begin{figure}[!h]
	\centering
	\includegraphics[width=1\textwidth]{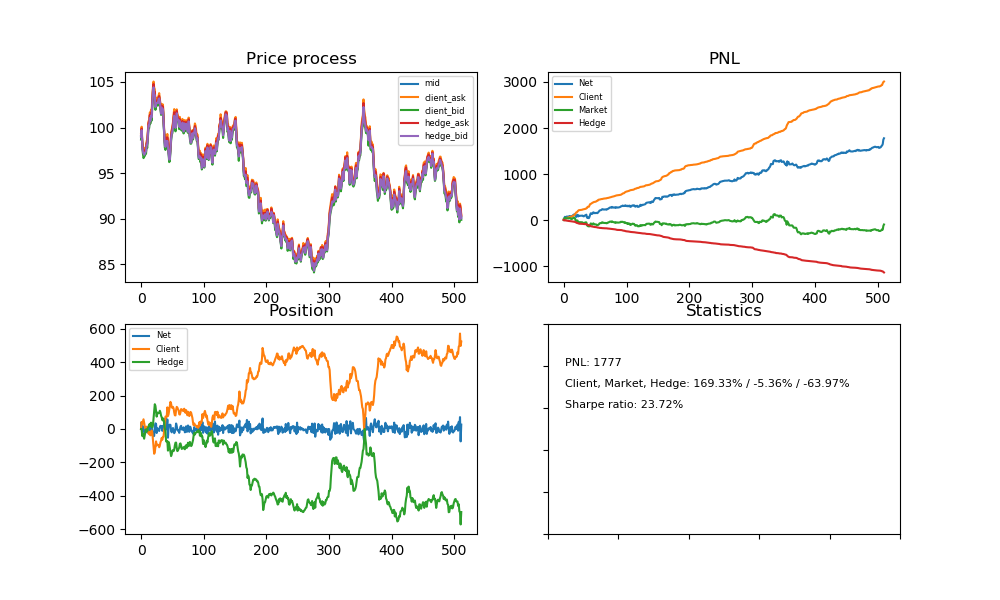}
	\caption{Dummy hedger dashboard}
	\label{Dummy hedger dashboard}
\end{figure}

It is of interest to set up the environment in such a way that the choice to offset the accrued client position with hedge trades should follow naturally from the observed reward. An obvious choice would be to use strategy profitability (PNL) as a basis for the reward function. Hence, we'd be using such parametrization of the environment that wrong hedging choices would result in adverse PNL. In particular, the ratio of mid process volatility $\sigma$ and stochastic spread multiplier $\nu$ is such that for a given realization of the mid price process the decision not to hedge quickly becomes apparent in strategy PNL.\\
To test this, let's use a dummy hedging strategy that on each step targets to offset the position accrued on the previous step. See \texttt{hedging\_env.py: def heuristic\_action}\\
To compare performance of such a strategy to an alternative not to hedge at all we will be using a dashboard that shows the price process, the net position alongside hedge and client positions, net PNL of the strategy and its constituent components: client PNL, market PNL and hedge PNL. With our dummy strategy on Figure \ref{Dummy hedger dashboard} it is seen that we are offsetting the outstanding client position with the opposite hedge position and are making the difference of client and hedge spreads. It is seen that the strategy looses on unfavorable spikes of market data between consecutive steps which is a result of position revaluation between steps. Compared to unhedged case this strategy exhibits low variance in PNL since market exposure is limited to only unhedged position increment (net client size flow) between steps.
\FloatBarrier
In the case of unhedged client position the variance of strategy PNL is driven by the combined variance of the simulated position and market data process. PNL becomes hostage to directional market moves resulting in worst Sharpe ratios.

\begin{figure}[!h]
	\centering
	\includegraphics[width=1\textwidth]{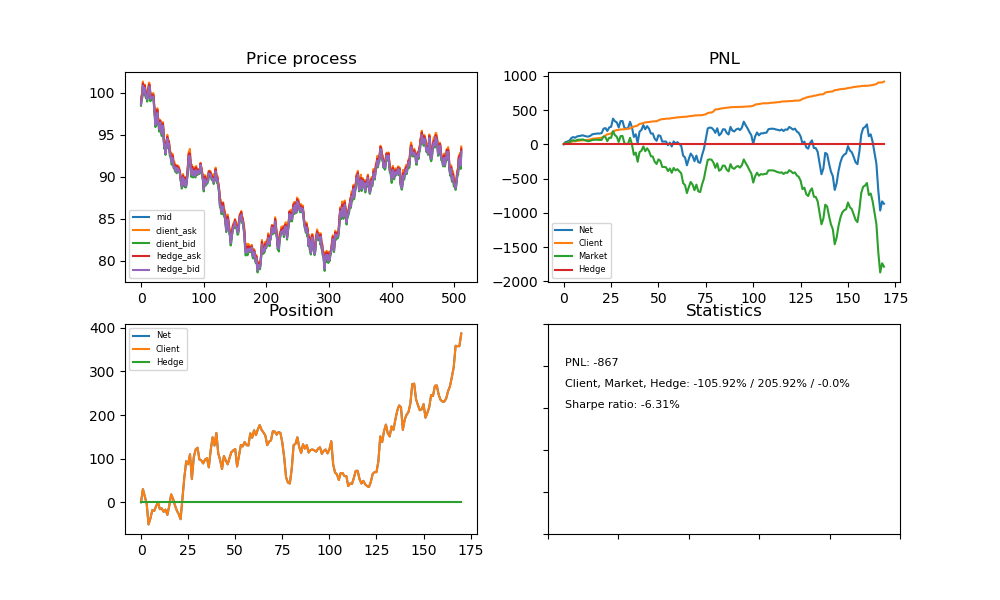}
	\caption{Unhedged client flow}
	\label{Unhedged client flow}
\end{figure}
\FloatBarrier

\section{Auto-hedging a single asset}
\label{Auto-hedging a single asset}
We want the learning agent to carry a position in time only if it is justified by the resulting PNL. As a matter of risk management we don't want the agent to take on excessive positions, so we need to set up a risk limit system. To achieve this goal a position penalty component is introduced to the reward function. This penalty is defined to be an exponential function of the position size, and is intended to make taking bigger positions increasingly more expensive for the agent in reward terms, forcing the agent to justify those by the resulting PNL. For small positions relative to $MaxPosLimit$ the charge would be relatively small but would increase exponentially when position nears or overshoots $MaxPosLimit$. We can see this penalty as a "spring" that would force the agent to bring the position back to zero unless having this position is profitable enough.

\begin{equation}
\label{Position penality}
Penalty = \gamma S_0  \left(e^{\frac{abs(Position)}{MaxPosLimit}} - 1 \right) MaxPosLimit
\end{equation}
where $\gamma$ constant determines how harshly the position penalty w.r.t $MaxPosLimit$ is enforced. To further discourage exploring unpromising trajectories the simulation is terminated when $MaxPositionLimit$ breach exceeds a certain multiple, and additionally penalized reward is recorded.\\ 
The following reward function is defined:
\begin{equation}
\label{Single autohedger reward function}
Reward_t = ClientPNL_t + HedgePNL_t + MarketPNL_t - Penalty_t
\end{equation}
Another technique to improve the convergence to the optimal policy is to limit the action space to a closed interval
\begin{equation}
\label{Hedge amount action}
a_i = hedgeAmount_i \in [-MaxHedgeAmount, MaxHedgeAmount]
\end{equation}
where $MaxHedgeAmount$ is defined to be of the same order as mean client trade size. This naturally limits the capacity of the learning agent to explore policies related to speculating on asset price movements and makes it focus on managing net position size instead. \\
Network architecture:
\begin{itemize}
	\item \textbf{action space}: agent action $a_i = HedgeSize \in [-MaxHedgeSize, MaxHedgeSize]$
	\item \textbf{observation space:} cumulative net position of client trades and the hedger up until and including the last step $NetPosition_i = \sum_{0}^{i}(NetClientTradeSize_j+HedgePositionSize_j) \in R$
	\item \textbf{network layers}: MLP of Linear (1) x ReLU (1024) x ReLU (1024) x Linear (1) layers (x batch\_size)
\end{itemize}
The agent successfully learns to hedge the client flow as it could be seen from rewards statistics and the simulation dashboard:
\FloatBarrier
\begin{figure}[!h]
	\centering
	\includegraphics[width=0.99\linewidth]{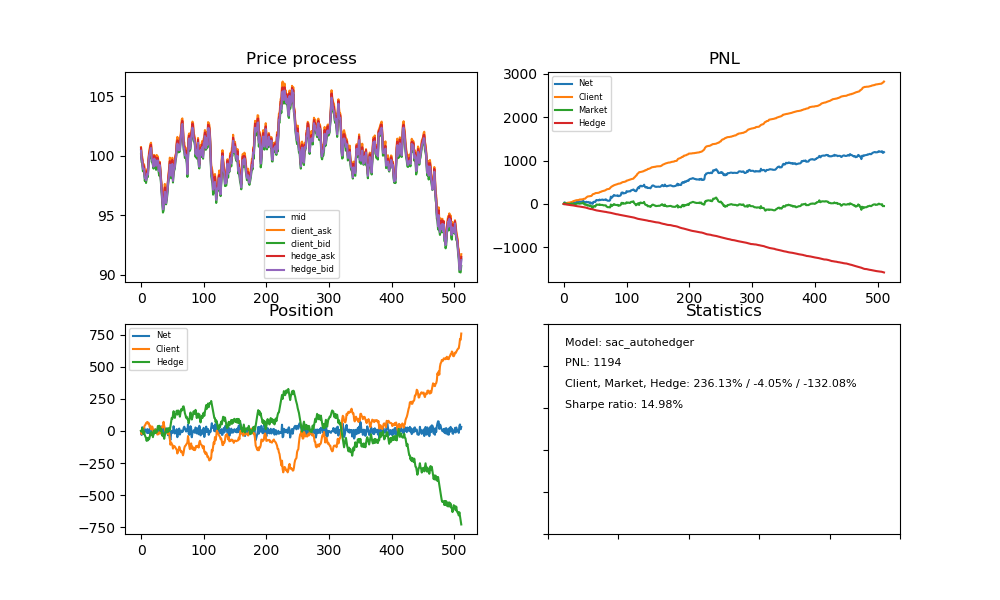}
	\caption{Autohegder single dashboard}
	\label{Autohedger single dashboard}
\end{figure}
\begin{figure}[!h]
	\centering
	\includegraphics[width=0.59\linewidth]{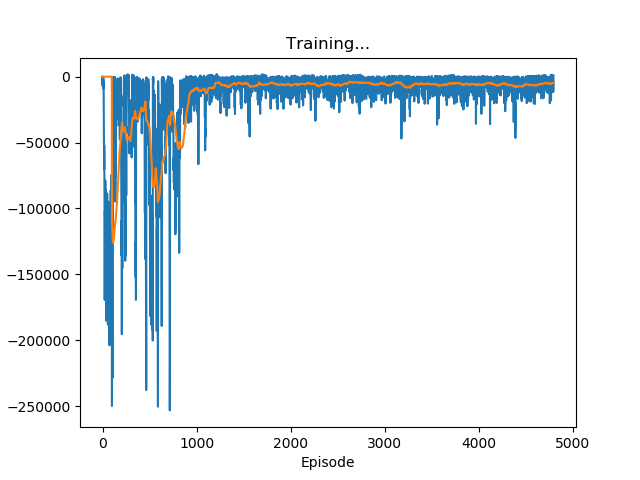}
	\caption{Autohedger single progress}
	\label{Autohedger single progress}
\end{figure}
\FloatBarrier
\begin{lstlisting}[language=bash, label=python-env-setup2,caption=Training the single asset autohedger]
cd stable-baselines3-autohedger-single/autohedger-single
python3 autohedger_ml.py
\end{lstlisting}
\section{Autohedging with skew}
\label{Autohedging with skew}
Let's now consider the setting where in addition to hedging on the market the agent can also apply skew i.e. adjust the offered price on one side to attract more offsetting client flow instead of hedging all of the position on the market. For simplicity we will assume that the offsetting client trade flow size is a linear function of skew (see Skewing model \ref{Skewing model explained}). \textbf{Architecture layout}:
\begin{itemize}
	\item \textbf{action space}: agent action is a pair of $\{HedgeSize, Skew\}$ rational numbers within a box bounded by $HedgeSize \in [-MaxHedgeSize, MaxHedgeSize]$ , $Skew \in [-1, 1]$
	\item \textbf{observation space:} net position
	\item \textbf{network layers}: MLP of Linear (1) x ReLU (2048) x ReLU (2048) x Linear (2) layers
\end{itemize}
It can be seen that the agent learns to make use of the skew in addition to utilizing hedge amounts to hedge the residual. This is useful as it allows us to raise the question of \nameref{Market price of risk}.
\begin{lstlisting}[language=bash, label=python-env-setup2,caption=Training the skew]
cd stable-baselines3-autohedger-single/autohedger-single
python3 autohedger_ml_skew.py
\end{lstlisting}
To gauge the performance of model with skew vs the model without skew it is useful to introduce a risk-weighted measure of reward. We'll be using a measure similar to Sharpe-ratio of returns, where a return of a single step $i$ is defined as $step\_return=PNL_{i} - PNL_{i-1}$:
\begin{equation}
\label{Sharpe ratio}
Sharpe\_ratio = \frac{E[returns\_array]}{stdev(returns\_array)} = \frac{\sum_{i=1}^{N\_steps}(PNL_{i}-PNL_{i-1})}{stdev(returns\_array)}
\end{equation}
To compare a non-skew and a skew model side-by-side those models are trained independently, saved, and then used on the same generated market data to compare the performance. 
\FloatBarrier
\begin{figure}[!h]
	\centering
	\includegraphics[width=1.1\linewidth]{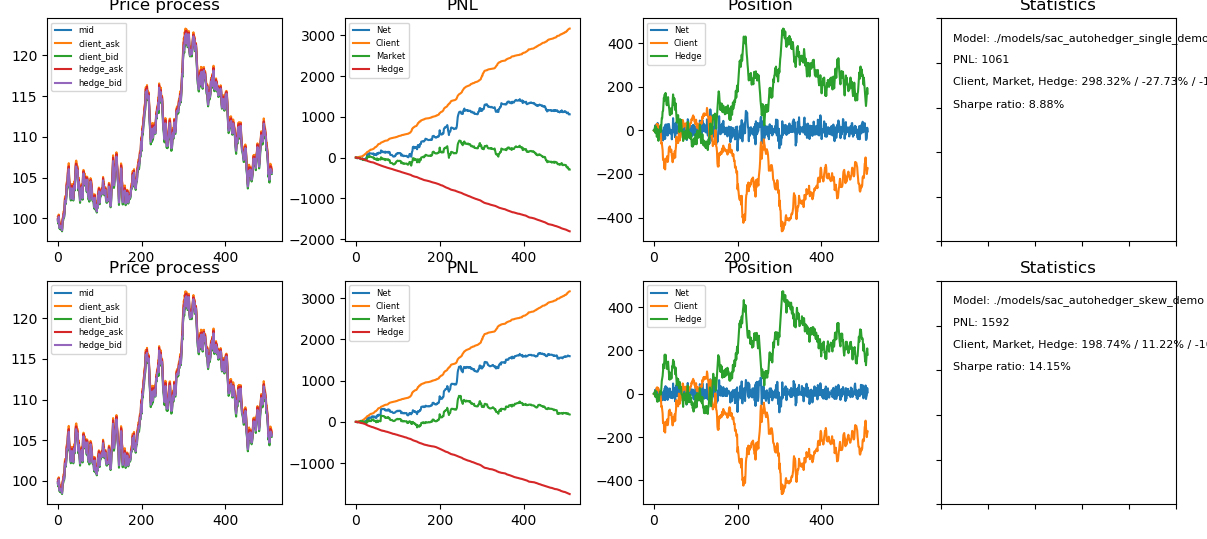}
	\caption{Models with skew and without}
	\label{Models with skew and without}
\end{figure}
\FloatBarrier
Depending on the magnitude of the chosen $\beta$ in the \nameref{Skewing model} it can be seen that skew-based model shows higher Sharpe ratios than a non-skew one. You can run the pre-trained models side-by-side on the same generated market data sets as per listing below:
\begin{lstlisting}[language=bash, label=python-env-setup2,caption=Comparison of models with skew and without]
cd stable-baselines3-autohedger-single/autohedger-single
python3 autohedger_ml_learned.py
\end{lstlisting}

\section{Market price of risk}
\label{Market price of risk}
Let's assume a setting where our market-making agent chooses which spreads to charge clients. It is of interest than to determine such minimum mean spread that would compensate the agent for the need of risk-managing a position in an asset resulting from the accumulation of the client trade flow. Such spread would include the cost of heding the position in the market as well as the premium for any non-Brownian behavior of the market (e.g. jumps, volatility clustering, drifts etc). Having those "hurdle" spreads calculated would allow the researcher to build the elasticity of demand spreading model on the top of them and to have a more clear idea of how to price the spreads being charged.\\
Within our market model the price process is a martingale, so the expected revaluation of a position between any two future timestamps is zero:
\begin{equation}
\label{martingale position}
Position_{t_1} * E[S_{t_2} - S_{t_1}] = 0 \;\;\; (t_2 \ge t_1 \ge 0)
\end{equation}
Within our PNL model this means that mean client spreads the agent should charge should be equal to mean hedge spreads paid on the exchange to offset the position. Note however that our agent does not know what those hedge spreads are and needs to work them out by interacting with the market model. \\
To achieve this goal the following reward function is defined:
\begin{equation}
\label{Market price of risk reward function}
Reward_t = max\left(-abs(ClientPNL_t + HedgePNL_t), 0\right) + MarketPNL_t - PositionPenalty_t
\end{equation}
This reward function makes the learning agent to come up with such client spreads that would offset the cost of hedging the position on the exchange. As before, we introduce a charge on carrying a directional position via $PositionPenalty$. As the agent learns, at first it incurs higher penalties due to exploring actions leading to larger positions. Once the agent learned to control the position, $PositionPenalty$ becomes a lesser factor letting the agent to bring client spreads in line with the cost of hedging.\\
We allow the learning agent to control client spreads via skew (see \ref{Skewing model explained}). In the context of this experiment we assume that client trade flow is not elastic to price changes so $\beta = 0$ in eq. \ref{Skewing model}.\\ \\
\textbf{Architecture layout}:
\begin{itemize}
	\item \textbf{action space}: agent action is a pair of $\{HedgeSize, Skew\}$ rational numbers within a box bounded by $HedgeSize \in [-MaxHedgeSize, MaxHedgeSize]$ , $Skew \in [-1, 1]$
	\item \textbf{observation space:} net position
	\item \textbf{network layers}: MLP of Linear (1) x ReLU (2048) x ReLU (2048) x Linear (2) layers
\end{itemize}
As a result of the learning process it is seen how rolling average spread charged by our agent ("maker" spread) converges onto the rolling average hedge spread ("taker" spread).
\begin{figure}[!h]
	\centering
	\includegraphics[width=0.49\linewidth]{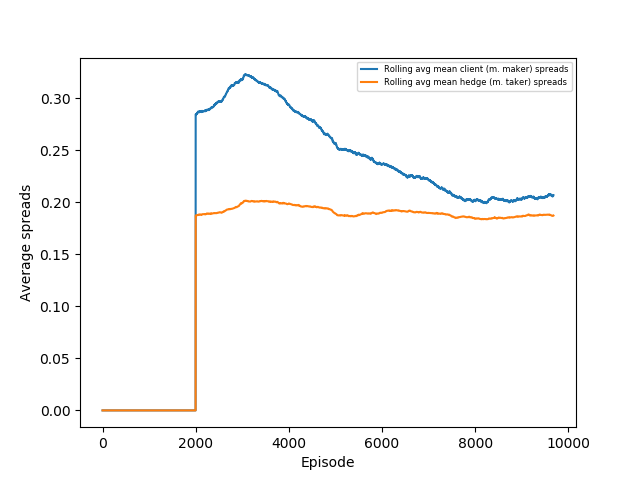}
	\includegraphics[width=0.49\linewidth]{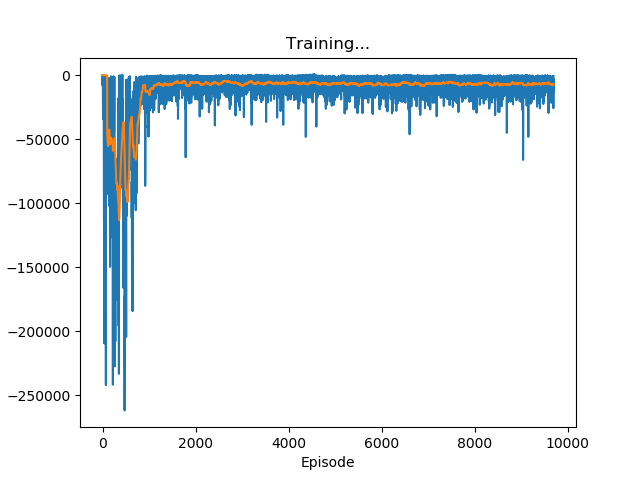}
	\caption{Discovering price of risk spreads}
	\label{Price of risk spreads}
\end{figure}
\FloatBarrier
On a sample PNL dashboard of the learned agent it is seen how the agent tries to keep flat net PNL, at the same time continuing to offset accrued client position with the hedger position.
\FloatBarrier
\begin{figure}[!h]
	\centering
	\includegraphics[width=1.0\linewidth]{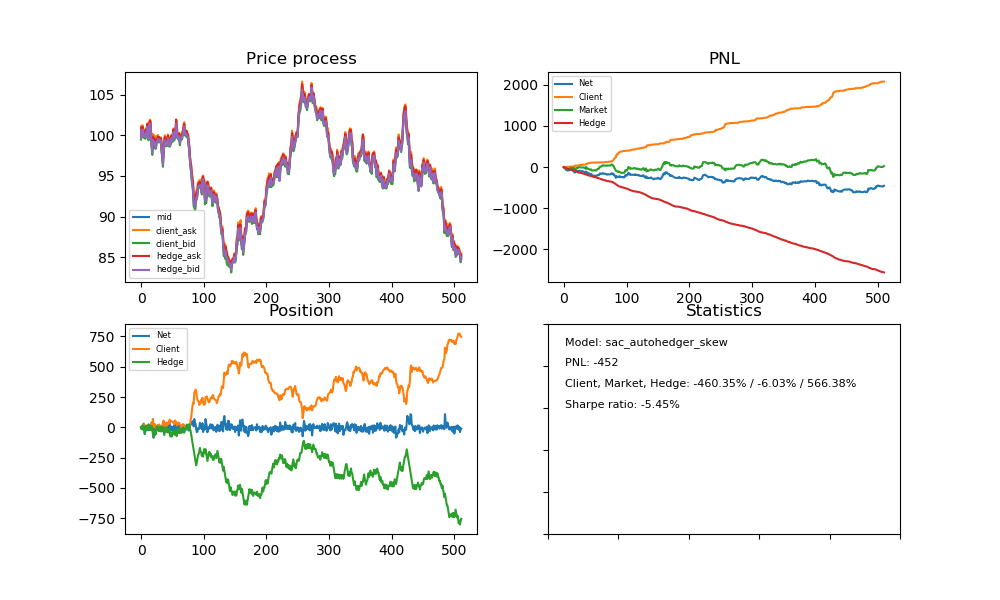}
	\caption{Price of risk dashboard}
	\label{Price of risk dashboard}
\end{figure}
\FloatBarrier
\begin{lstlisting}[language=bash, label=python-env-setup2,caption=Training Price of risk model]
cd stable-baselines3-autohedger-single/autohedger-single
python3 autohedger_ml_price_of_risk.py
\end{lstlisting}

\section{Modeling portfolios}
Let's assume we have two assets $S^1$ and $S^2$ which log increments \ref{Asset price process 1} are correlated as in \ref{Correlated} with correlation coefficient $\rho$. For simplicity let's assume that log-increments have the same flat volatility $\sigma$, so that the resulting log-normal price processes would have the same variance characteristics. Each of $S^1$ and $S^2$ processes would drive its own \nameref{TradeFlowModel} and \nameref{Stochastic spread model}. Naturally, each of those processes would imply its own Poisson intensity of arriving client trade sizes, as well as net imbalance of the arriving client trade flow. "Portfolio" is formed by blending these asset prices and net client trade size flows \ref{Net trade size for a step} with a fixed weight $w$:
\begin{equation}
\label{Portfolio prices}
S^{blended}_{mid,bid,ask} = w S^1_{mid,bid,ask} + (1 - w) S^2_{mid,bid,ask}
\end{equation}
\begin{equation}
\label{Portfolio sizes}
TradeSize_{net}^{blended} = w TradeSize_{net}^1 + (1 - w) TradeSize_{net}^2
\end{equation}
As a result, the reinforcement learning agent accrues a position which gets incremented with the blended size and the blended price. The agent does not know what asset weights are being used ($w$ and $1-w$), neither does it know the correlation coefficient $\rho$. The agent needs to learn the optimal hedging strategy using hedge amounts for $S^1$ and $S^2$. \\
At any step the net position ("portfolio") value is defined as the sum of values of the blended client position and its hedges:
\begin{equation}
\label{Portfolio value}
Portfolio\_value(t) = S^{blended}_{mid}(t) Position_{Client}^{blended}(t)  + S^1_{mid}(t) Position_{Hedge}^{S^1}(t) + S^2_{mid}(t) Position_{Hedge}^{S^2}(t) 
\end{equation}
The goal of the learning agent would be to maximize policy PNL whilst keeping the position value in check. The action space is now a $R$-valued square which bounds the sampled hedge amounts $S^1$ and $S^2$. As before, we will be using a penalty function to facilitate risk-management by the learning agent. However, a straightforward application of the exponential penalty function (see single-asset penalty function \ref{Position penality}) to $Portfolio\_value$ would not be as effective. The reason is that by having the two-dimensional action space we are now allowing $HedgeAmount^{S^1}$ and $HedgeAmount^{S^2}$ to offset each other, thus introducing an additional degree of freedom to the action space. This shows up as a lower net portfolio value and contradicts the convex definition of the blended price $\ref{Portfolio prices}$. This could be remediated in a few ways. \\
The first approach of making the actions convex is to change the parametrization of the action space:
\begin{equation}
\label{Convex parametrization S1}
HedgeAmount^{S^1} = w_{action} * hedge\_amount_{action}
\end{equation}   
\begin{equation}
\label{Convex parametrization S2}
HedgeAmount^{S^2} = (1 - w_{action}) * hedge\_amount_{action}
\end{equation}
Here actual actions are the weight $w$ and $hedge\_amount$, and the respective $S^1$ and $S^2$ hedge amounts are being calculated in a convex manner from those.\\
A better approach would be to leave it for the learning agent to reason out but to introduce an additional "over-hedge" penalty for when an individual hedge value overshoots the value of the blended client position or has the same sign (so leverages instead of offsetting):
\[
Overhegdge=\begin{cases}
abs(hedge\_value + client\_pos\_value) \;\;\; \text{ hedge overshoots or same sign}\\
0 \;\;\;\;\;\;\;\;\;\;\;\; \text{ other cases}
\end{cases}
\]
Then $Overhedge$ is used as a part of the overall penalty function, where $\phi$ constant determines penalty tolerance to overhedge and $\gamma$ determines the tolerance to overall portfolio value closing on $MaxPosLimit$ or overshooting it:
\begin{equation}
\label{Position penality portfolio}
Penalty = \gamma S_0  \left(e^{\frac{abs(Portfolio\_value) + \phi{abs(Overhedge)}}{S_0 MaxPosLimit}} - 1 \right) MaxPosLimit
\end{equation}
Note that three positions are accrued over the simulation trajectory: the blended client position, $S^1$ hedge position and $S^2$ hedge position, where hedge positions are supposed to be offsetting the blended client position. For each of those positions at each step we are tracking spread PNL (paid or received) as well as market revaluation PNL.\\ 
Portfolio PNL is defined as trajectory PNL consisting of blended client position PNL (client spread PNL and client position market revaluation) and the sum of $S^1$ and $S^2$ hedge positions PNL (hedge spread PNL and hedge position market revaluation). Reward function is  defined as portfolio PNL minus the penalty function:
\begin{equation}
\label{Portfolio agent reward}
Reward\_portfolio = BlendedClientPosition\_PNL + HedgePosition^{S^1}\_PNL + HedgePosition^{S^2}\_PNL - Penalty
\end{equation}
\textbf{Architecture layout}:
\begin{itemize}
	\item \textbf{action space}: agent action is a pair of $\{HedgeSize\_S^1, HedgeSize\_S^2\}$ rational numbers within a square box bounded by $HedgeSize \in [-MaxHedgeSize, MaxHedgeSize]$ 
	\item \textbf{observation space:} $S^1$ hedger position value $S^1_{mid} Position_{Hedge}^{S^1}$, $S^2$ hedger position value $S^2_{mid} Position_{Hedge}^{S^2}$ and net $Portfolio\_value(t)$ (see $\ref{Portfolio value}$) 
	\item \textbf{network layers}: MLP of Linear (input: 3) x ReLU (8000) x ReLU (8000) x Linear (output: 2) layers
\end{itemize}

Given the initial correlation $\rho=0$, the agent is seen to learn to offset the blended incoming client size flow with the hedge positions. The agent seems to be able to deduce the implied weighting of assets within the blended client trade flow, however, more work needs to be done on improving stability on farther epochs. One can find sample dashboards for portfolio learning agents in \nameref{Portfolio hedging dashboards}.
\begin{lstlisting}[language=bash, label=python-env-setup2,caption=Training portfolio hedge model]
cd stable-baselines3-autohedger-portfolio/autohedger-portfolio
python3 autohedger_ml_portfolio.py
\end{lstlisting}
A slower, more stable version:
\begin{lstlisting}[language=bash, label=python-env-setup2,caption=Training portfolio hedge model - slow]
cd stable-baselines3-autohedger-portfolio/autohedger-portfolio
python3 autohedger_ml_portfolio_slow.py
\end{lstlisting}

\section{Conclusion}
Techniques explored in this paper could be used as a baseline for setting up a position management framework within an automated market-making system. More importantly, however, is that it could be used as a way of blending alpha-generation and position management into one reinforcement learning system. Let's assume that we have a signal generating unit that is listening to the real-life price process as well as other information sources like order book, news feeds etc. Then the signal it generates could be plugged into our reinforcement learning agent. If (as a result of the learning process) our agent finds this signal to be beneficial to the reward function, it would automatically learn to adjust its hedging decisions in line with that. Bridging signal-generation capacity with reinforcement-learning inventory management poses an interesting area of research.\\
Automating portfolio management decisions by letting the agent infer correlations and weights while learning also proves to be a challenging area to be explored. In a more realistic setting for the portfolio management task we could define a risk-weighted reward function (e.g. based on Sharpe ratio $\ref{Sharpe ratio}$) and let the agent pick from tens or hundreds of assets which would translate into the same dimensionality for observation and action spaces and would increase the complexity of stochatic policy search problem.\\
Even in the current set-up more work needs to be done to increase learning stability and avoid intermittent oscillations of $HedgeSize$ action between $-MaxHedgeSize$ and $MaxHedgeSize$ between consecutive steps on farther epochs. This is most likely related to diminished scale of rewards on farther epochs compared to large scale of penalties incurred while starting to train. Methods like normalization of rewards and entropy regularization could be used for this purpose, which would boil down to fine-tuning the entropy maximization parametrization within SAC.

\section{Appendix}
\subsection{Project repository}
Project code is available under \textbf{https://github.com/bakalex/autohedger}
\subsection{Physical environment}
Modeling was performed on Amazon EC2 g4dn.2xlarge machine with the use of Deep Learning AMI (Ubuntu 16.04) Version 30.0 (ami-02379288a3b4cbe7b). PyTorch 1.5 CUDA 10.1 is activated with:
\begin{lstlisting}[language=bash, label=python-env-setup,caption=Setting up PyTorch environment on EC2 AMI]
source activate pytorch_latest_p36
\end{lstlisting}
g4dn.2xlarge allowed for research on MLPs as large as 10000 x 10000 x 8000 (x default batch size), but typically nets of no more than 8000 x 8000 units were used. \\
Open AI baseline installed with:
\begin{lstlisting}[language=bash, label=python-env-setup2,caption=Installing Open AI baseline]
cd stable-baselines3
pip install .
\end{lstlisting}
For the convenience of reproducing the results raised in this paper, a version of Open AI is supplied alongside the project code. It is a choice of the researcher to use it or to try a more recent version.

\subsection{SAC algo}
\label{SAC learning loop}
Soft actor-critic algorithm as listed in \href{https://spinningup.openai.com/en/latest/algorithms/sac.html}{Open AI Spinning Up}. A more succinct version is available in \cite{haarnoja2018soft1}.

\FloatBarrier
\begin{figure}[!h]
	\centering
	\includegraphics[width=0.8\linewidth]{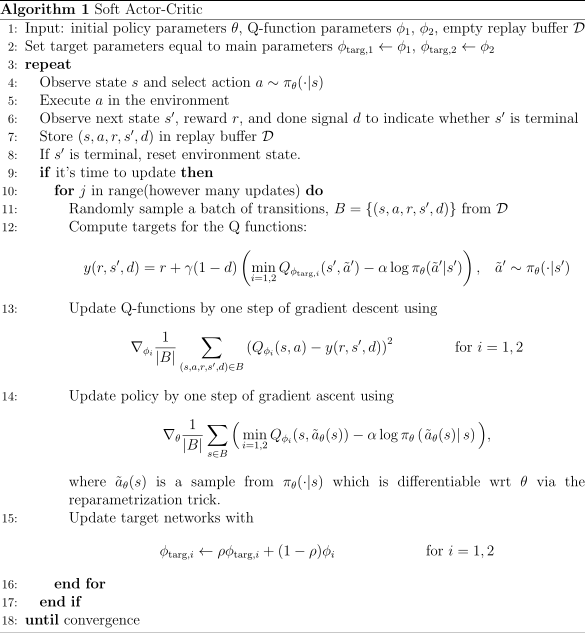}
	\caption{Soft actor-critic RL algorithm}
	\label{Soft actor-critic RL algorithm}
\end{figure}
\FloatBarrier

\subsection{Portfolio hedging dashboards}
\label{Portfolio hedging dashboards}
\FloatBarrier
\begin{figure}[!h]
	\centering
	\includegraphics[width=0.9\linewidth]{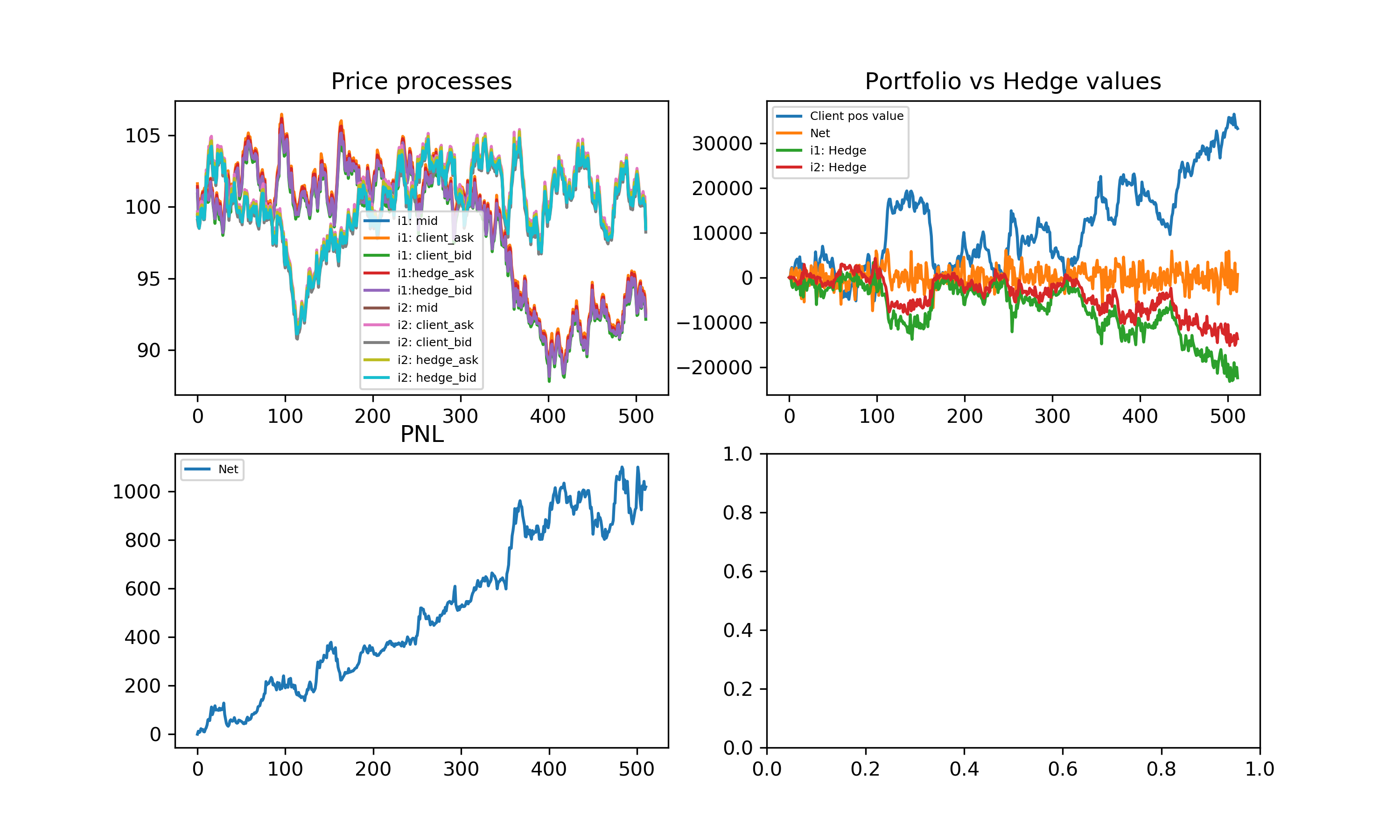}
	\caption{Learning to hedge portfolios, w=0.5}
	\label{Learning to hedge portfolios 1}
\end{figure}
\begin{figure}[!h]
	\centering
	\includegraphics[width=0.9\linewidth]{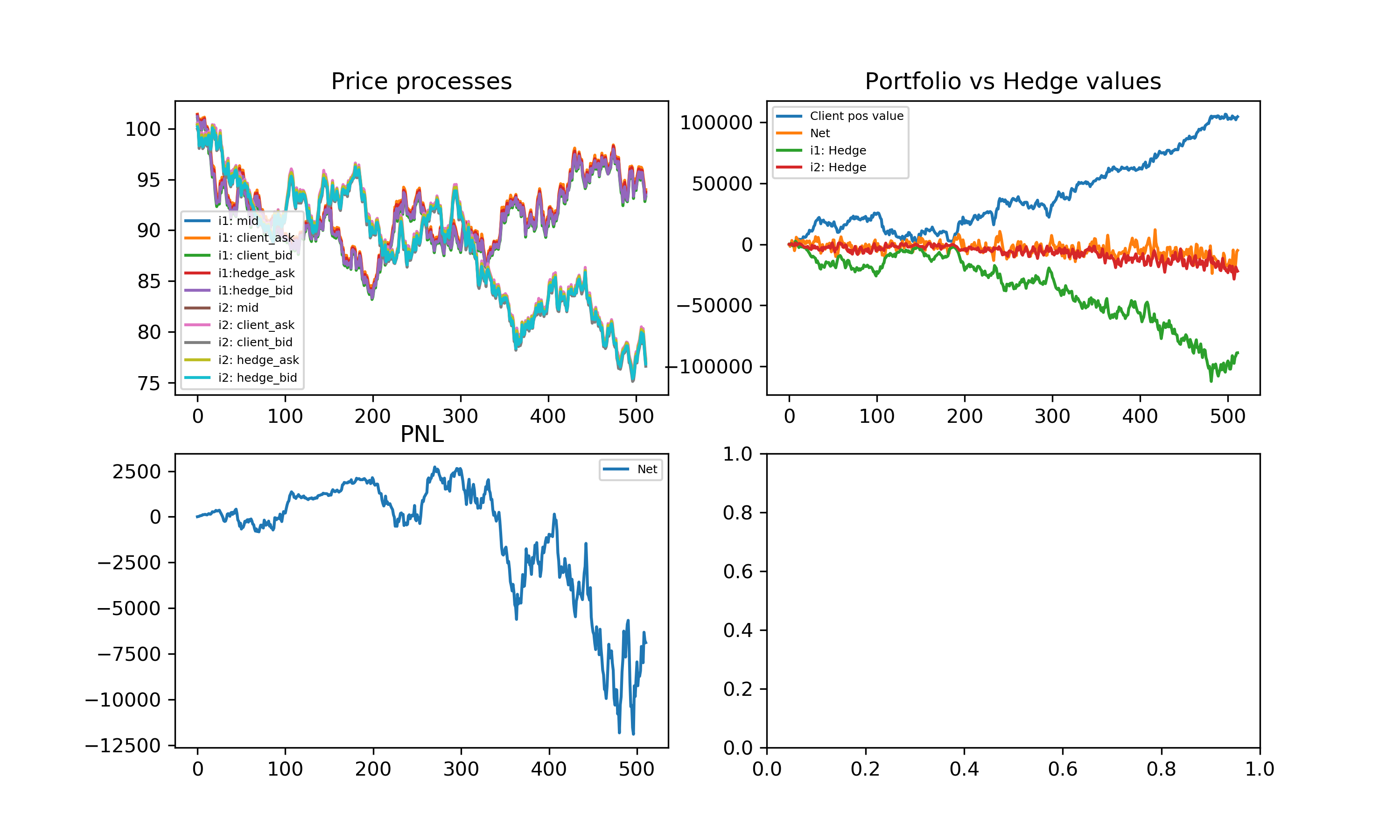}
	\caption{Learning to hedge portfolios, w=0.0}
	\label{Learning to hedge portfolios 2}
\end{figure}
\FloatBarrier
\nocite{*}
\bibliography{myrefs2}

\begin{thebibliography}{1}

\bibitem{abdolmaleki2018maximum}
Abbas Abdolmaleki, Jost~Tobias Springenberg, Yuval Tassa, Remi Munos, Nicolas
  Heess, and Martin Riedmiller.
\newblock Maximum a posteriori policy optimisation, 2018.

\bibitem{fox2015taming}
Roy Fox, Ari Pakman, and Naftali Tishby.
\newblock Taming the noise in reinforcement learning via soft updates, 2015.

\bibitem{fujimoto2018addressing}
Scott Fujimoto, Herke van Hoof, and David Meger.
\newblock Addressing function approximation error in actor-critic methods,
  2018.

\bibitem{haarnoja2018learning}
Tuomas Haarnoja, Sehoon Ha, Aurick Zhou, Jie Tan, George Tucker, and Sergey
  Levine.
\newblock Learning to walk via deep reinforcement learning, 2018.

\bibitem{haarnoja2018soft}
Tuomas Haarnoja, Aurick Zhou, Pieter Abbeel, and Sergey Levine.
\newblock Soft actor-critic: Off-policy maximum entropy deep reinforcement
  learning with a stochastic actor, 2018.

\bibitem{haarnoja2018soft1}
Tuomas Haarnoja, Aurick Zhou, Kristian Hartikainen, George Tucker, Sehoon Ha,
  Jie Tan, Vikash Kumar, Henry Zhu, Abhishek Gupta, Pieter Abbeel, and Sergey
  Levine.
\newblock Soft actor-critic algorithms and applications, 2018.

\bibitem{henderson2017deep}
Peter Henderson, Riashat Islam, Philip Bachman, Joelle Pineau, Doina Precup,
  and David Meger.
\newblock Deep reinforcement learning that matters, 2017.

\end{thebibliography}
\end{document}